# Habitat fragmentation reshapes genomic footprints of selection in a forest herb

Van Daele Frederik, Honnay Olivier, De Kort Hanne

Department of Biology, KU Leuven, Leuven, Belgium

## Abstract

Understanding the combined effects of climate change and habitat fragmentation on the adaptive potential of plant populations is essential for devising effective conservation strategies. This is particularly important where mating system variation impacts the evolutionary consequences of habitat fragmentation. Here we aimed to reveal how habitat fragmentation and climate adaptation jointly influence the evolutionary trajectories in *Primula elatior*, a heterostylous self-incompatible and dispersal-constrained forest herb. We quantified the genomic variation and degree of herkogamy, a floral trait reducing self-pollination, across 60 geographically paired populations of *Primula elatior* across Europe, each pair featuring contrasting levels of habitat fragmentation. Our findings revealed a large and unique set of adaptive outliers in more fragmented landscapes, compared to high-connectivity ones, despite the geographic proximity of the sampling pairs. This suggests elevated selective pressures in fragmented habitats, mirrored by a reduced adaptive potential to cope with climate change. Finally, a minority of genetic variants associated with herkogamy were influenced by current levels of habitat fragmentation and population size, potentially signalling early indicators of evolutionary mating system changes in response to pollinator limitation. Because evolutionary trajectories and adaptive potential are expected to be increasingly affected by habitat fragmentation, our findings underscore the importance of considering both habitat fragmentation and climate adaptation in conservation research and planning.

## Introduction

Many species are faced by both climate and habitat fragmentation, making their evolutionary responses to these factors highly interrelated. Climate-induced selective pressures significantly shape local adaptation in plant populations by driving evolutionary changes in specific traits and processes (Franks & Hoffmann, 2012). Climate factors introduce stressors such as temperature extremes, drought, and fluctuating resource availability, which can lead to genetic changes or stimulate epigenetic responses in plant genomes (Crisp et al., 2016; Savolainen et al., 2013). These genetic processes form the basis for natural selection, facilitating the long-term establishment of adaptive traits as they add to a repository of genetic diversity that equips plant populations to more effectively respond to rapidly changing climatic conditions (Torres-Martínez et al., 2019; Yeaman & Whitlock, 2011). However, habitat fragmentation can potentially disturb these adaptations, through influencing critical plant traits like growth rates, phenological timing, and physiological adaptations, including drought resilience (Franks et al., 2014; Nicotra et al., 2010). Yet, to what extent habitat fragmentation alters the evolutionary trajectories associated with climate remains poorly known.

The efficacy of local adaptation depends on the interaction between natural selection and gene flow. Natural selection drives the local fixation of beneficial alleles, while gene flow introduces new genetic material into populations (Kawecki & Ebert, 2004). In heterogeneous landscapes, this balance is further complicated by the relation between demography and habitat patch size, patch isolation, and varying degrees of connectivity between patches (Hanski et al., 2011; Rees et al., 2000). Fragmentation of habitats often leads to restricted gene flow and smaller effective population sizes, heightening the risk of inbreeding, genetic drift, and genetic bottlenecks, in turn potentially disrupting local adaptations and their adaptive potential (Aguilar et al., 2006; Leimu et al., 2010). The fitness in these populations can be further compromised through a reduced efficacy of purifying selection, which primarily acts to remove deleterious mutations (Habel & Zachos, 2012). Populations in highly fragmented habitats can also experience intensified selective pressures from factors like edge effects and interspecific interactions, thereby compromising or reinforcing climate adaptation (Cheptou et al., 2017; Legrand et al., 2017). On the other hand, gene swamping in contiguous landscapes poses its own risk by potentially diluting locally advantageous alleles (Lienert, 2004; Polechová & Barton, 2015). As a result, mild habitat fragmentation can be expected to facilitate natural selection especially if remaining populations maintain high levels of standing genetic diversity (Yeaman & Otto, 2011). In addition, unique genomic signatures of climate adaptation can arise in fragmented settings through antagonistic pleiotropy, whereby a single gene can have opposing effects on multiple traits impacting fitness, such as survival and reproductive success (Hamilton et al., 2015; Lascoux et al., 2016). Thus, habitat fragmentation can profoundly influence the adaptive potential of populations, by altering genetic diversity and influencing evolutionary processes critical for coping with environmental changes (Dubois & Cheptou, 2017).

Habitat fragmentation may have a particularly pronounced impact on the adaptive potential of outcrossing species, such as angiosperms with a herkogamous mating system (the spatial separation of anthers and stigmas). A balanced morph ratio is pivotal for these species, as it promotes disassortative pollen transfer and outcrossing. However, habitat loss can increase the random chance of losing certain mating types in a population. This becomes especially problematic when there is not enough genetic exchange between remaining populations to maintain the balance that negative frequency-dependent selection typically provides (Charlesworth & Willis, 2009; Kirk & Freeland, 2011). Furthermore, in fragmented habitats, the pollinator availability is often reduced, further disrupting plant-pollinator interactions, with accumulating evidence that this can drive evolutionary changes in plant reproductive traits. One significant trend is the increased prevalence of autonomous selfing in fragmented habitats to ensure reproduction under pollen limitation (Jacquemyn et al., 2012). This evolutionary shift is often associated with morphological changes, such as the reduction in the level of herkogamy when plant-pollinator networks are disrupted (Brys & Jacquemyn, 2015; Eckert et al., 2010). In such cases, the diallelic S-locus supergene, which controls reciprocal herkogamy and determines the length of thrum and pin morphologies, becomes critically important as it serves as a fundamental genetic mechanism for maintaining self-incompatibility (Li et al., 2007, 2015). Moreover, habitat fragmentation has the potential to exert selective pressures on this S-locus supergene. These pressures can influence important pollination parameters like pollen limitation, conspecific pollen proportions, and mate availability, potentially tipping the scales of negative frequency-dependent selection that usually maintains S-allele diversity (Li et al., 2015). As a result of these pressures, balancing selection at the S-locus can be disrupted, leading to a shift from obligate outcrossing to increased selfing, mediated specifically by changes in the S-locus (Bawa et al., 2011; Encinas-Viso et al., 2020). Such genetic shifts often herald the replacement of heterostylous morphs with self-compatible homostyles (Yuan et al., 2023). While these genetic and morphological shifts can ensure short-term reproductive success in fragmented and pollinator-poor environments, they may diminish genetic diversity and long-term adaptive potential (Zhang et al., 2021). The adaptive benefits conferred by the S-locus, such as the maintenance of genetic variation and reciprocal herkogamy, become compromised, posing risks for reduced genetic recombination and heightened inbreeding depression (Porcher & Lande, 2005).

Focusing on *Primula elatior* subsp. elatior, a species with specific reproductive and dispersal challenges that can serve as a model for other dispersal-limited species, this study seeks to disentangle the multifaceted interactions between genomic adaptation to climate, habitat fragmentation, and reproductive isolation mechanisms. Specifically, using genomic, morphological, and ecological data, we aim to evaluate to what extent i) genomic signatures of adaptation vary between fragmented and connected landscapes; ii) habitat fragmentation affects the adaptive potential of natural populations under climate change; and iii) evolutionary patterns of herkogamy align with climate and/or habitat fragmentation. By exploring these interactions, this study aims to provide vital insights into the adaptive capacity of species in fragmented and rapidly changing environments, offering implications for effective conservation strategies.

## 1. Methods

### 2.1 Study species and sample collection

*Primula elatior* subsp. elatior, a herb species typical of moist European oak and oak-hornbeam woodlands, served as our model for studying how habitat fragmentation affects climate adaptation, reproductive isolation, and the adaptive potential of dispersal-limited plant species. The choice of our model species was driven by reproductive self-incompatibility and seed dispersal limitations, representative for many broadleaf forest herb species in Europe. Reciprocal herkogamy, a distinct floral trait involving morphological differences in male and female floral organs, plays a significant role in the species' reproductive success and gene flow (Keller et al., 2016). Furthermore, the species is limited in its seed dispersal capabilities due to the absence of specific adaptations like structures facilitating wind or animal dispersal (Jacquemyn et al., 2002).

To explore the role of habitat fragmentation and climate adaptation in shaping genome-wide variation and reproductive traits of *P. elatior*, we employed a strategic sampling approach, collecting leaf and flower samples from 60 paired populations across the distribution range (Van Daele et al., 2022; Van Daele et al., 2023). Specifically, a fragmented and contiguous habitat were selected at each of 30 latitudes, based on four specific criteria. These criteria included a Mean Patch Proximity Index (a measure reflecting both isolation and fragmentation of a patch) of less than 115 for fragmented patches and greater than 250 for connected patches (Bender et al., 2003; Gustafson & Parker, 1994), a Mean Nearest Neighbour distance (representing the average shortest distances to deciduous forest patches) of less than 2000 meters for connected patches and more than 2000 meters for fragmented patches (Elkie et al., 1999), a population size of fewer than 400 for fragmented patches and greater than 600 for connected patches (Jacquemyn et al., 2001), and an elevation below 200 meters (Taylor & Woodell, 2008).

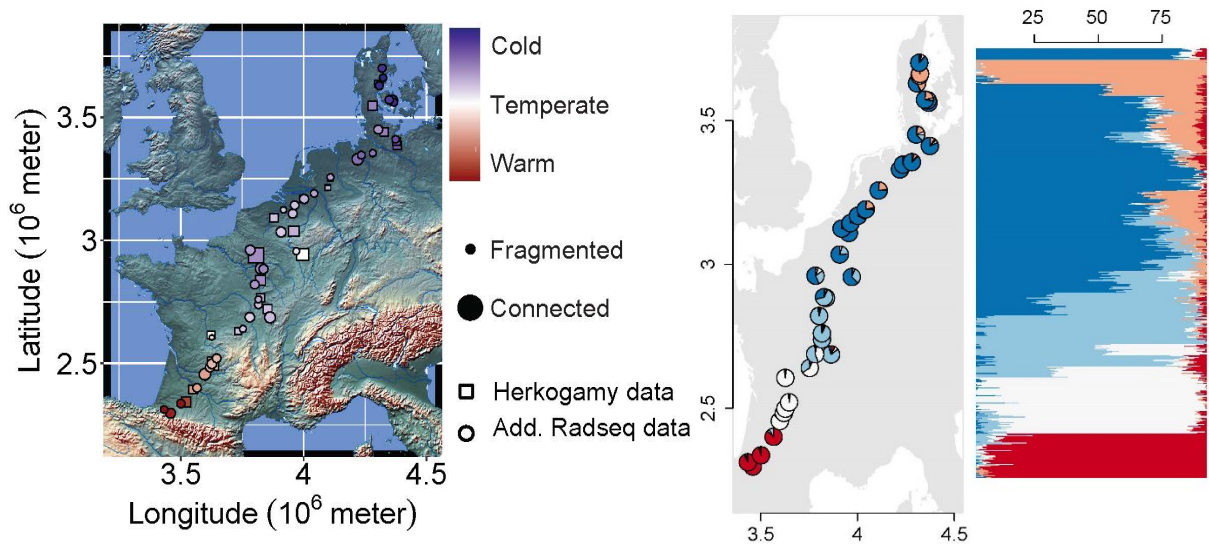

*Figure 1: The locations of the sample collection is depicted in panel A. Herkogamy data was collected in all 60 locations. The radseq data was collected in 38 of these locations. A geographic depiction of admixture proportions and individual ancestry coefficients for radseq data within* Primula elatior *populations is presented in panel B, spanning from southern France to north Denmark (left). The spatial extent of the distribution range across Europe is denoted in meters × $10^6$ (Coordinate system: LAEA89) along the longitudinal (x) and latitudinal axes (y). The admixture proportions are displayed as percentages for each individual within the populations studied.*

In each population, leaf samples from thirteen individual plants were harvested, with a minimum distance of 100 metres between individuals. These samples (n=462) were stored in silica gel, and genomic DNA was subsequently extracted using a Plant DNA Extraction Kit (Norgen Biotek, 2015). To examine the stability of self-incompatibility traits, we also measured herkogamy for three flowers from each of 2116 individual plants across the 60 populations. In addition, we assessed the population sizes across all 60 populations, with sizes ranging from 7 to over 1000 flowering individuals (counted up to 1000).

## 2.2 Read assembly and pre-processing

Our primary genomic analysis utilised a restriction site-associated DNA marker sequencing approach (RadSeq) on leaf samples from 38 populations, which included 30 fragmented and 8 re-sequenced contiguous populations. RADseq data were generated from 12-13 individuals in each of these populations (with the exception of one fragmented population that only had seven individuals), resulting in a total of 462 samples.

DNA concentration was quantified with PicoGreen using the Picofluor method (BioSystems, 2007), 2007). Genomic DNA was then digested using a dual restriction enzyme combination procedure (BtgI/TaqI) and separate libraries were assembled with dual-indexed genotyping by sequencing (Shao et al., 2020). All libraries were combined into a single pool and sequenced on a single lane of a NovaSeq 6000 system with the S4 flow cell type (96 samples per lane), producing $2 \times 150$ bp with Illumina NovaSeq Reagent Kits according to the manufacturer's instructions (Illumina, 2020). The `Burrows-Wheeler Aligner (BWA-MEM) v. 0.7.17´ (Li, 2013; Li & Durbin, 2009) was used on CentOS7 to align reads to a novo *Primula veris* reference genome (NCBI assembly code: ASM78844v1; Nowak et al., 2015). This resulted in 5'822'200 ± 280 (se) raw reads. Fastq files were then evenly subsampled down to a maximum of 7'000'000 reads per sample, which resulted in an average of 5'605'138 ± 176 (se) reads. Reads were then removed if no 5' cut site was found, and if the sequencing adapter was removed from the 3' end but no 3' cut site was found, or if the trimmed read was too short (Garbe, 2022). The bam alignment files were then sorted and indexed with `Samtools v. 1.16.1´ (Li et al., 2009). Regions of bam files with more than 500 reads were downsampled to a depth of 500 reads using VariantBam v. b797b78 (Wala et al., 2016). `Freebayes v. 1.3.6´ (Garrison & Marth, 2012) was used to call variants jointly across all samples while selecting the best 4 alleles, based on the ranked sum of supporting quality scores, and a minimum coverage of 960. The raw VCF file generated by Freebayes was filtered using `VcfFilter v. 0.2´ (SASC, 2018) to remove the lowest quality variants with a Phred score smaller than 20, which indicates a 99% probability that there is a variant at the site. This resulted in 114'214 SNPs with 1.66 % missing data. We used `dartR v. 2.7.2´ (Mijangos et al., 2022) to further filter SNP quality based on a minimum allele frequency (MAF) of 0.05, SNPs that were not monomorphic in all selected samples, and loci that were not biallelic. We found a positive relationship between read depth and the SNP error rate and therefore we removed the lowest 15% of the read depth distribution (read depth >= 7.3) to avoid low-confidence SNP calls (O'Leary et al., 2018). SNPs with the highest 5% of the read depth distribution (read depth >= 118.7) were removed to avoid overclustered loci. SNPs with at least 1 sequencing error in the duplicate samples were removed. To optimize data quality, any SNPs with remaining missing data (call rate = 1) were removed. Finally, we filtered based on the linkage disequilibrium with a maximum distance of 200 base pairs and an $R^2$ threshold of 0.1, where SNPs with the highest call rate were retained. This resulted in a database of 10'309 SNPs for further analysis.

## 2.3 Explanatory variables

For the assessment of habitat fragmentation, we utilised a combination of landscape metrics to assess broadleaf forest habitat, derived from Copernicus Land Monitoring Service (2018) with a spatial resolution of 100m. Metrics were calculated using the R packages 'raster v.3.4-13' (Hijmans, 2019) and 'landscapemetrics v.1.5.4' (Hesselbarth et al., 2019). A 2000m buffer radius was established around each patch centroid, marking the boundary for primary seed and pollen dispersal. The landscape context of habitat patches was assessed with 11 metrics, indicative for the size, shape and isolation of individual habitats. Specifically, we calculated mean patch area, aggregation index (He et al., 2001), cohesion (Schumaker, 1996), division (Jaeger, 2000), mean gyrate (Keitt et al., 1997), number of patches, and class percentage of the landscape. These metrics were reduced in dimensionality with principal component analysis using a varimax rotation (Revelle, 2022). The first principal component, capturing 58.4% of the total variance with a sum of squared loadings of 5.84, was retained for further analysis. Notably, lower values of this component were associated with increased levels of habitat fragmentation (Appendix 1). In order to account for density-dependent effects, we also integrated population size alongside landscape fragmentation.

To capture the climatic variability across the 60 studied habitats, we extracted two specific bioclimatic variables from WorldClim version 2.1 (Hijmans, 2019): the maximum temperature of the warmest month and the precipitation seasonality (coefficient of variation). Among all bioclimate variables, these two variables were found to drive climate adaptation in the species according to an earlier study (Van Daele et al., 2022).

## 2.4 Population genomic structure

To identify outlier loci significantly deviating from background genetic variation, we employed a dual approach. First, we utilised 'PCAdapt v.4.3.3' (Luu et al., 2017) with 20 principal components in order to detect the optimal number of principal components (K=3) capturing background population structure based on the scree plot of eigenvalues (Appendix 2). The first three components explained 16.8% of the total variance, thus providing a reasonable representation of the underlying population structure for the subsequent outlier analysis. Second, we used 'BayeScan v.2.1' (Beaumont & Balding, 2004) on the dataset purged of linkage disequilibrium. This Bayesian method estimates the posterior probability of a locus being under selection, thereby minimising false positives across diverse demographic scenarios (Foll & Gaggiotti, 2008). Given the dimensionality of our dataset, a prior odds of 10 was employed in BayeScan, congruent with observed outlier proportions in preceding studies (Van Daele et al., 2022). The analysis included 100 pilot runs, each 5000 iterations long, followed by a main run of 100,000 iterations with a 50,000 iteration burn-in period (Lotterhos & Whitlock, 2014). False Discovery Rates (FDRs) were evaluated based on q-values and were assessed at thresholds ranging from 0.1% to 50%, with a precision set at 0.1%. The locus-specific effect, indicated by the alpha parameter in Bayescan, serves as a marker for deviations from neutrality: a positive alpha suggests diversifying selection, while a negative one hints at balancing or purifying selection (Foll & Gaggiotti, 2008). Diversifying selection highlights adaptive divergence between populations, while balancing selection reveals evolutionary processes upholding intra-population diversity, such as frequency-dependent selection, which may play a key role in herkogamous species (Barrett, 2019).

To accurately evaluate ancestry across the range and control for demographic influences in our environmental association analysis, we employed sparse non-negative matrix factorization (sNMF) algorithms. Utilising the 'LEA v.3.4.0' R package (Frichot et al., 2014), we explored genetic clusters within a range of 1 to 25, identifying the optimal number of clusters as five based on the cross-entropy criterion. This optimal cluster number, K=5, harmonises with ecological observations and previous determined ancestry analysis from closely related datasets (Van Daele et al., 2022) . Subsequent k-means clustering categorised populations according to their genetic composition, delineating distinct spatial genetic clusters throughout the distribution. These clusters, hereafter referred to as Demography, were instrumental for incorporating background genetic structure in the environmental association analysis.

## 2.5 Environmental association analysis

### 2.5.1 Disentangling drivers of local adaptation

To evaluate to what extent genetic variation is associated with climate, habitat fragmentation, herkogamy, and population size while accounting for the neutral genetic structure influenced by the spatial arrangement of our populations, we employed a redundancy analysis (RDA) using the R package 'Vegan v.2.5-7' (Capblancq et al., 2018; Orsini et al., 2013). The formula for our RDA model was:

$$Genotype \sim Frag.PC1 + Pop.Size + Temp. + Prec. + \bar{x}\,herkogamy + Cond.(Demography)$$

In this model, *"Genotype"* denotes the biallelic genotypic matrix object where each allele of a SNP is coded as 0, 1, or 2, signifying the number of copies of the reference allele an individual possesses (Mijangos et al., 2022). The *"Frag. PC1"* represents the first principal component of the habitat fragmentation variables, pinpointing genetic outliers influenced by habitat fragmentation. *"Pop. Size"* encapsulates the size of each population, which may drive selection at loci underpinning mating system and gene flow across the genome (Barrett, 2019; Cheptou & Massol, 2009). "*Temp.*" and "*Prec.*" represent the maximum temperature of the warmest month and precipitation seasonality, respectively, targeting genetic outliers associated with climatic variations (Fick & Hijmans, 2017). The $\bar{x}$ herkogamy represents the average herkogamy, determined from measurements of three flowers per plant. Lastly, Cond.(Demography) integrates spatial genetic clusters as conditional variables, aiming to separate the effects of spatial organisation from the influences of climate and habitat fragmentation, ensuring the observed patterns aren't merely due to spatial genetic structure (encoded as factor).

To pinpoint SNPs with strong associations to the explanatory variables in our RDA model, we extracted the loadings and identified SNPs that deviated significantly from the mean using a threshold of 2.5 standard deviations (Borcard et al., 2011; Legendre & Legendre, 2012). Focusing on the top four (K=4) principal components (based on the scree plot elbow), we assessed the correlations between these candidate SNPs and our predictors, honing in on the maximum correlation to discern the most robust associations (Legendre et al., 2011). Two biplots of SNP and site scores were subsequently generated to visually elucidate the relationships between the candidate SNPs, the populations, and the explanatory variables (Holderegger et al., 2015).

By linking the diversifying and balancing distinctions made by BayeScan (based on the alpha parameter) to the outliers detected by other methods (PCAdapt and RDA), we aimed to maintain a consistent evolutionary framework across different analytical approaches. This ensures that the comparative insights derived are grounded in the same evolutionary understanding, allowing for a more coherent interpretation of adaptive processes across the range. The overlap of diversifying and balancing outliers (as detected by Bayescan) between BayeScan, PCAdapt, and the full RDA was then evaluated using `ggVennDiagram v. 1.2.2´ (Gao, 2022).

### 2.5.2 Impact of habitat fragmentation on outlier distributions

To evaluate how habitat fragmentation affects signatures of climate adaptation and outliers (aim i), we first categorised populations in two classes (fragmented/connected) using the median of the first principal component of habitat fragmentation variables (2.3 explanatory variables). To ensure a balanced and geographically representative spread of both fragmentation classes across the range, we adopted a distance-based approach, selecting an equal number of populations from each fragmentation status (Appendix S1). This approach not only offers a precise understanding of habitat fragmentation's genomic implications on local adaptation but also provides a nuanced perspective on how fragmentation might influence diversifying and balancing outliers in local adaptation. The comparative analysis of diversifying and balancing outliers was carried out across the complete dataset, and the subsets of connected and fragmented habitats, employing both BayeScan and PCAdapt. The overlaps were then visualised using using `ggVennDiagram v. 1.2.2´ (Gao, 2022).

To evaluate the impact of habitat fragmentation on the climate adaptive capacity of *Primula elatior* plant populations, we compared the output of three population-level Redundancy Analyses on (i) the full dataset, (ii) the dataset of fragmented habitat, and (iii) the dataset of connected habitat (from here on called climate RDA):

$$MAF \sim Temp. + Prec. + Cond.(Demography)$$

In these RDA models, the minor allele frequencies (*"MAF"*) of the genotype matrix were calculated and Hellinger-transformed. Specifically, Hellinger transformation involves taking the square root of the relative abundance values of minor alleles within each population. This Euclidean-based measure allows for more accurate comparisons between populations by accounting for differences in allele frequencies. To prevent overfitting, population-level allele frequencies were used as response matrix because the predictors only capture population-level variation. To pinpoint climate-related SNPs in the three datasets (full, fragmented, and connected), we analysed their loadings as *per* 2.5.1. Focusing on the top two principal components, we evaluated their maximum correlation with the climate drivers to identify the strongest associations. By comparing the adaptive climate outliers detected in fragmented and connected habitat with the full dataset using `ggVennDiagram v. 1.2.2´ (Gao, 2022), we could evaluate how habitat fragmentation disrupts the adaptive genomic architecture of *Primula elatior*.

To evaluate the genetic diversity available for climate adaptation (aim ii), we also calculated the adjusted expected heterozygosity ($He_{adj}$ = He * n loci / {n loci + n.invariant}) of outliers detected with the climate RDA in the full dataset using DartR v. 2.7.2 (Mijangos et al., 2022). Here, the genetic diversity ($He_{adj}$) was calculated for neutral SNP frequencies (after the removal of adaptive outliers), as well as the climate outliers. The four genetic diversity ($He_{adj}$) metrics of each of the 38 populations derived from the full RDA climate analysis where categorized based on the median fragmentation PC1 (factor) and compared with linear mixed-effects models (lmer) using `lme4 v. 1.1-31´ (Bates et al., 2015):

$$He_{adj} \sim Frag.\ group + (1\ |\ Latitude)$$

In this context, the rounded latitude values were calculated using the LAEA89 coordinate system and resulting metrics in metres as unit were divided by $10^6$ and used as a random effect to control for regional influences. This coarse-grained latitude metric enabled the aggregation of geographically proximate sites into broader regional units, simplifying spatial complexity and allowing for the evaluation of regional-scale patterns in genetic diversity and adaptation. Predicted means and confidence intervals were derived from the lmer using `ggeffects v. 1.1.4´, conditional and marginal $R^2$ values were calculated using performance v. 0.10.1 (Lüdecke et al., 2021), and chi² and p-values were calculated using a type 3 ANOVA using `car v. 3.1-1´ (Fox & Weisberg, 2019).

### 2.5.3 Investigating evolutionary drivers of herkogamy

To investigate the evolutionary consequences of antagonistic selection pressures acting on herkogamy (aim iii), we contrasted the RDA which included all explanatory variables (chapter 2.5.1) to a herkogamy-only RDA:

$$Genotype \sim \bar{x}\,herkogamy + Cond.(Demography)$$

The comparison of this herkogamy-only model to the full RDA model (2.5.1) served to elucidate to what extent the genomic basis of herkogamy is independent of genomic signatures of habitat fragmentation and climatic, or rather algins with these selective pressures. Similar to the previous models, we extracted the loadings of SNPs that deviated from the mean by at least 2.5 standard deviations, focusing on the top principal component.

## 2. Results

### 2.1. Genomic architecture and local adaptation in fragmented habitats

We observed a considerable number of decisive outliers under diversifying and balancing selection (Fig. 2A, B). BayeScan identified a substantial number of outliers under both diversifying (9.96%) and balancing (17.3%) selection, which were notably higher compared to those identified by PCAdapt for diversifying (5.74%) and balancing (0.3%) selection.

Both BayeScan and PCAdapt analyses revealed distinct sets of outliers that were specific to either fragmented or connected habitats (Fig. 2C-F). The PCAdapt analysis in particular found a greater number of diversifying outliers in fragmented as compared to connected habitats, potentially indicating the presence of unique or more intense selection pressures in fragmented settings. Signatures of balancing selection were most abundant in the full dataset, and less pronounced in the connected and fragmented subsets.

Our redundancy analysis (RDA) revealed that habitat fragmentation, population size, temperature, precipitation seasonality, and mean herkogamy all were significantly associated with the genomic architecture of *Primula elatior*. The RDA model accounted for 6.35% of the total variance (adj. $R^2$) in the genomic dataset. The first two adaptive RDA axes were mainly associated with temperature (RDA1, $r = 0.083$) and precipitation seasonality (RDA2, $r = 0.059$) where $r$ denotes the correlation coefficient with the respective environmental variable.. Habitat fragmentation and population size were correlated with RDA3 ($r = 0.053$ and $r = -0.058$, resp.). Herkogamy was not significantly associated with the major RDA axes, but contributed slightly to the genetic variation captured by RDA5 ($r = -0.052$), which only explained 2.98% of all constrained genomic variation. The RDA outliers only partially overlapped with the diversifying outliers identified by Bayescan and PCadapt (Fig. 3C), suggesting that Bayescan and PCadapt identified adaptive signals not associated with our RDA predictors. Furthermore, overlap between RDA, bayescan, and PCAdapt was almost completely absent for balancing outliers (Fig. 3D).

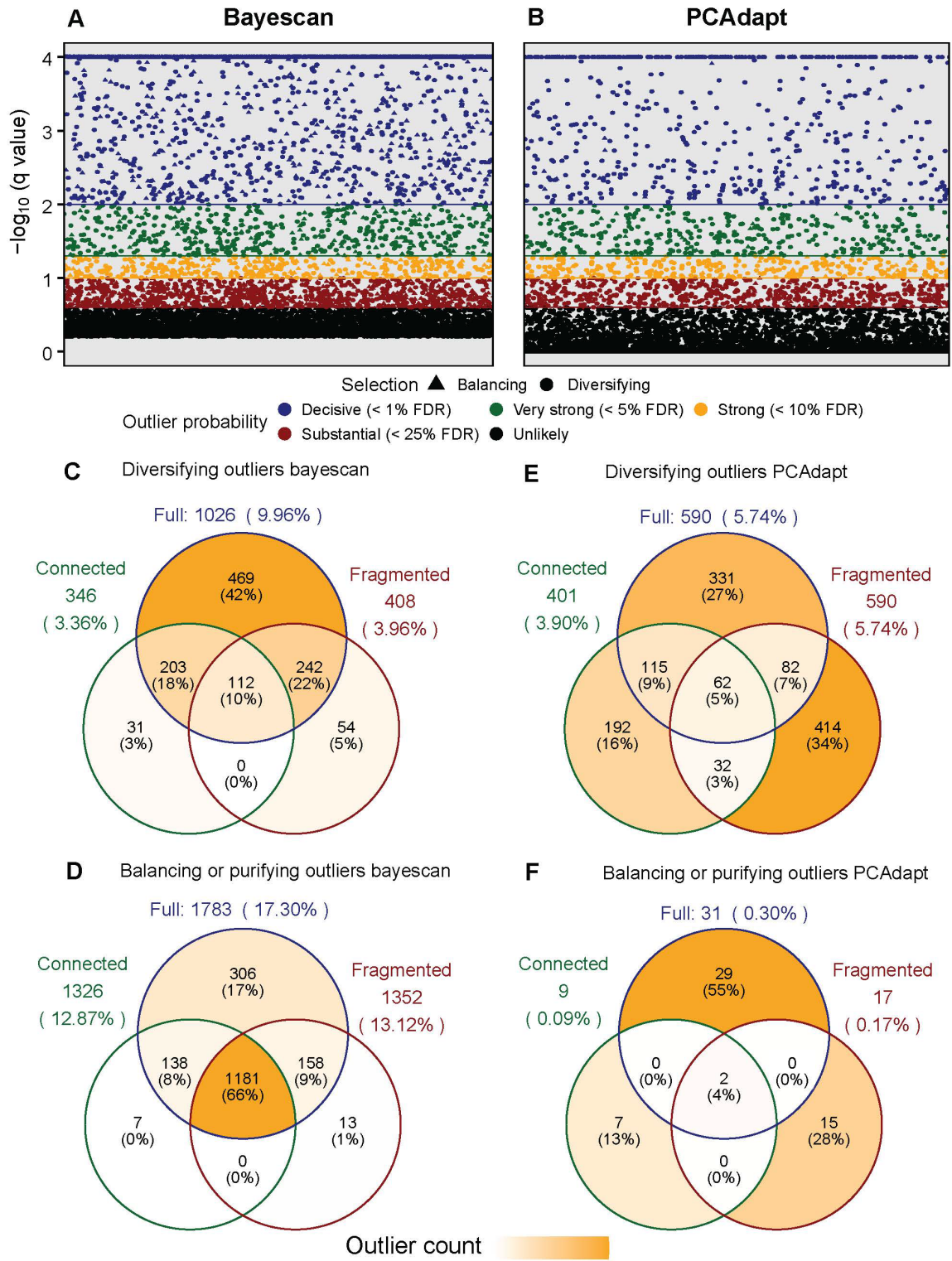

*Figure 2: Panels A and B present the Manhattan plots for BayeScan and PCAdapt results respectively, where SNP outlier significance (-log₁₀ q value) is denoted on the y-axis and outlier probability thresholds determined by colours. Panels C to F depict Venn diagrams illustrating the overlap of significant SNPs under diversifying, balancing or purifying selection, derived from BayeScan (C; D) and PCAdapt results (E; F), where only decisive outliers (< 1% FDR) were considered. Within the Venn diagrams, the percentage sums across sections total 100%, whereas counts outside the diagrams signify the aggregate of decisive outliers for specific subsets (full, connected, fragmented), with percentages indicating their proportion in relation to the complete SNP-dataset.*

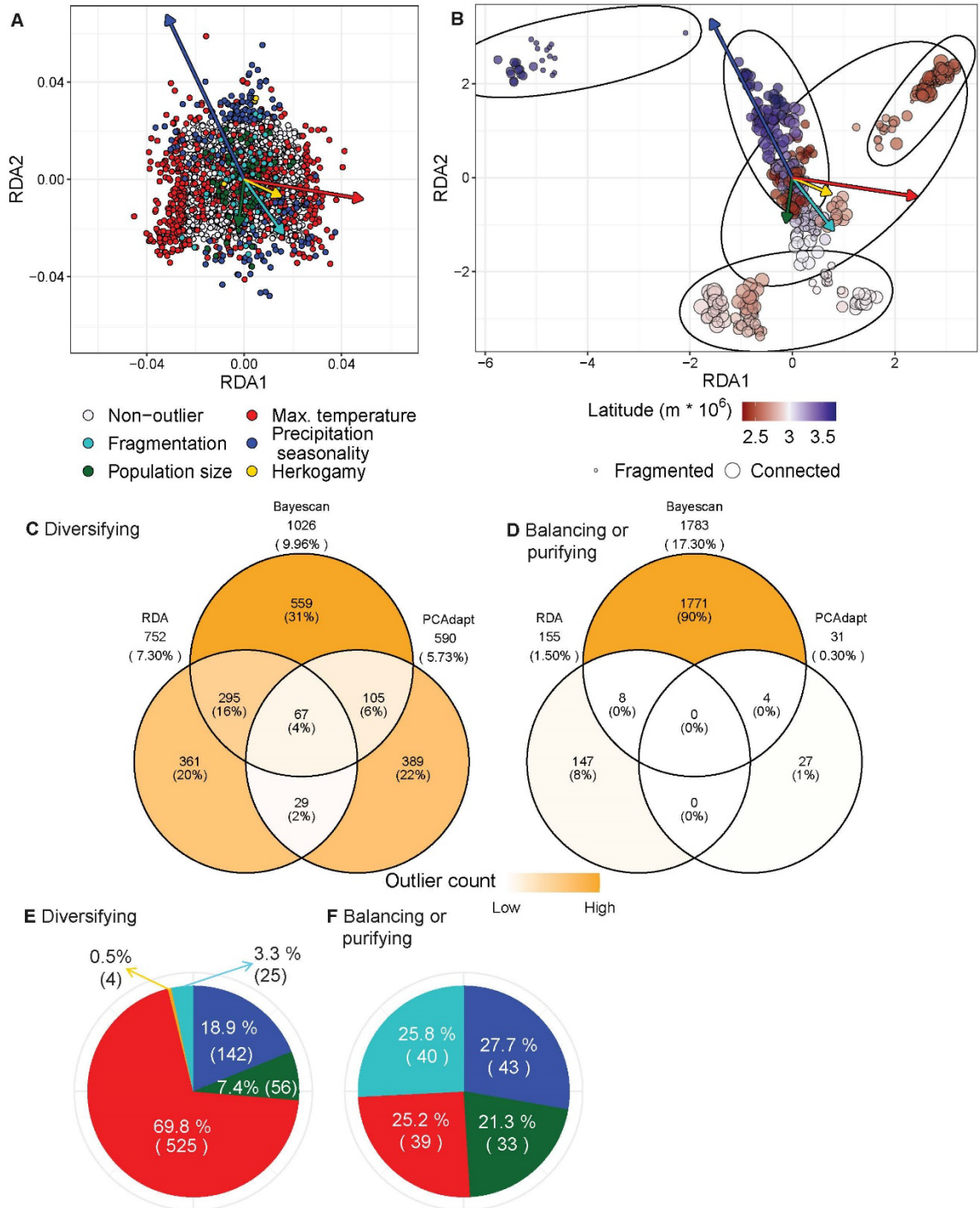

*Figure 3: Visual representation of SNP and site biplots of the full RDA model (2.5.1 Disentangling drivers of local adaptation), and comparisons between outlier detection methods. Panel A displays a biplot of SNPs, underpinned by the explanatory roles of local adaptation drivers and herkogamy in shaping outlier patterns. RDA1, with an eigenvalue of 100.55, accounted for 52.04% of the explained variance, while RDA2, having an eigenvalue of 46.98, contributed an additional 24.32%. Panel B delineates the biplot of site scores, capturing the relationship between species sampling sites and their associations with latitude (indicated by colour) and habitat fragmentation (signified by size). Panel C displays a Venn diagram comparing diversifying outliers as identified by RDA, Bayescan, and pcadapt. Percentages within the diagram represent the proportion of outliers specific to each method, while percentages outside denote the proportion of outliers in relation to the total number of SNPs. Panel D operates similarly but focuses on balancing outliers. Panel E displays the diversifying outlier proportions concerning RDA's fixed environmental drivers, while Panel F outlines the balancing outlier comparisons in relation to these fixed drivers (colours depict the fixed drivers and relate to the colour legend of the SNP biplot outliers).*

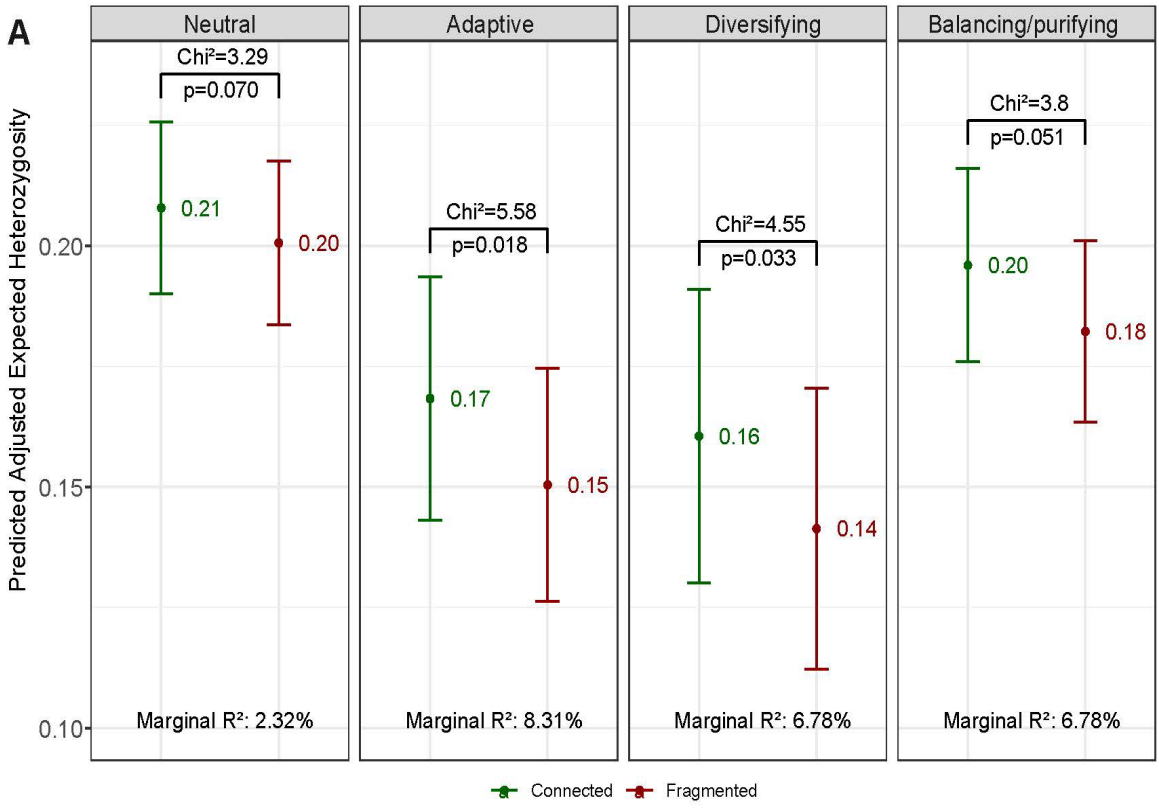
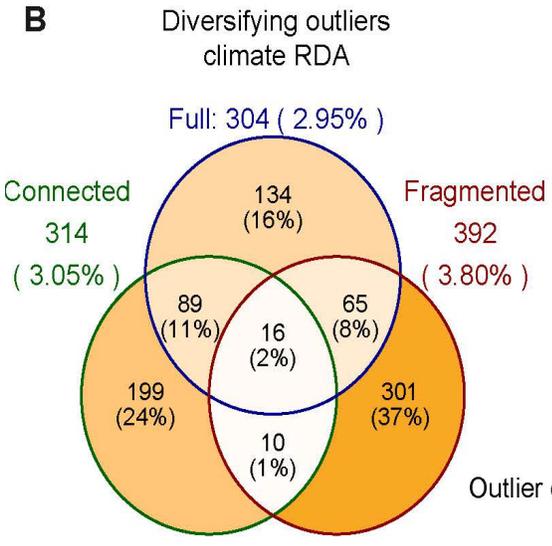
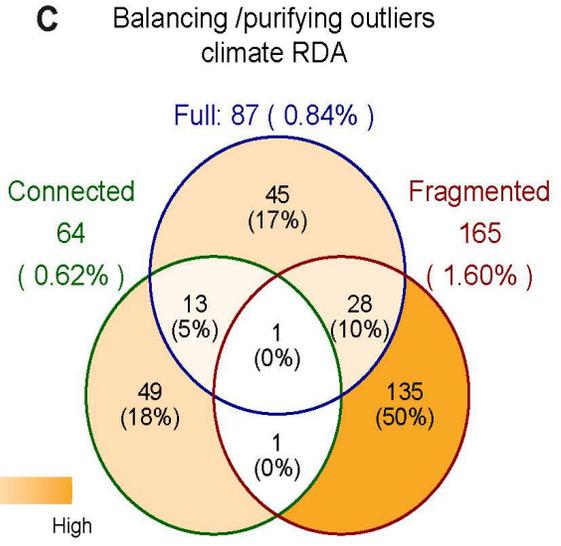

*Figure 4: Effects of habitat fragmentation on climate adaptation in* Primula elatior. *For Panel A, adjusted heterozygosity levels (He) were modelled using a linear mixed-effects model (lmer) with habitat fragmentation groups (as determined by the median value of the fragmentation PC1 metric) as the fixed effect and rounded latitude (LAEA89 in meters / $10^6$) as a random intercept Distinct lmer models were used to calculate the predicted means and 95% confidence intervals of genetic diversity metrics derived from the climate RDA: neutral genetic diversity (after removing adaptive outliers), adaptive outliers (encompassing both diversifying and balancing climatic outliers), diversifying climate outliers, and balancingor purifying climate outliers. Panels B and C present Venn diagrams illustrating the overlap of climate outliers for diversifying and balancing or purifying selection, respectively, between the full climate model and fragmented and connected habitat subsets. Numbers within each Venn diagram segment indicate the count of unique climate outliers specific to each condition as well as those shared between the two. Within the Venn diagrams, the percentage sums across sections total 100%, whereas counts outside the diagrams signify the aggregate of outliers for specific subsets (full, connected, fragmented), with percentages indicating their proportion in relation to the complete SNP-dataset.*

Taking the first four RDA axes into account, the majority of outliers were predominantly associated with climatic variables (564 temperature and 185 precipitation outliers, resp.; Fig 3E, F). Habitat fragmentation and population size were each significantly associated with 65 outliers. Furthermore, four outliers were predominantly associated with mean herkogamy (Fig. 3E), which showed a directional alignment with temperature and habitat fragmentation (Fig. 3B, 3E), potentially suggesting that these SNPs represent environment-dependent mating system variation.

## 2.2. Impact of habitat fragmentation on adaptive potential and herkogamy

Climate-related genetic diversity was slightly, but significantly lower in fragmented as compared to connected habitats (Fig. 4A). Neutral genetic diversity, on the other hand, was not significantly affected by habitat fragmentation. Similar to BayeScan and PCAdapt, the climate RDA revealed that fragmented populations harboured a greater number of outliers, potentially indicating more intense selection pressure in fragmented settings (Fig. 4B, 4C).

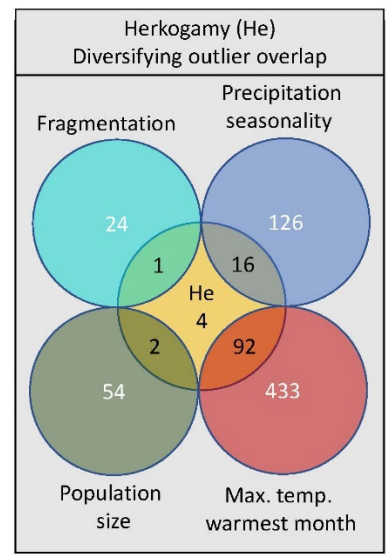

Figure 5: Overlap between herkogamy detected in the full model (4) and outliers detected in the herkogamy-only model.

When the genetic variants associated with herkogamy were compared between a herkogamy-only model and the full model, herkogamy outliers (N=242) appeared to be predominantly correlated to climate (N=92 and 16 for temperature and precipitation seasonality, resp.). Furthermore, one diversifying outlier was predominantly related to fragmentation in the full model and two to the population size.

## 3. Discussion

As the climate changes at an unprecedented rate and forest landscapes remain fragmented across Europe and the world, understanding how these factors determine the adaptive capacity of dispersal-limited species is vital for effective conservation. Habitat fragmentation led to a large and distinct panel of outliers under both diversifying and balancing selection in *P. elatior*, indicating altered selective pressures in fragmented landscapes. This was associated with a reduced adaptive potential for climate-related genetic variation, suggesting that populations subject to habitat fragmentation are more vulnerable to future climate change than populations in connected landscapes. The genetic variants influencing herkogamy were found to be primarily correlated with climate and, to a lesser extent, with population size and habitat fragmentation. Thus, while herkogamy can be suggested to align with habitat fragmentation to some extent, other variables have hitherto left a more profound genomic footprint on herkogamy. We argue that the reductions in adaptive potential and intensified selection pressures already associated with habitat fragmentation may presage further loss of evolutionary potential, because long-living and clonal species such as *P. elatior* respond slowly to environmental shifts (Vellend et al., 2006).

### 3.1. The effects of habitat fragmentation on the adaptive architecture

Genetic outliers were predominantly influenced by extreme temperatures and variation in seasonal precipitation. This is in line with previous studies identifying strong climate-driven diversifying selection across the range of *Primula elatior* (Van Daele et al., 2022). Our population genomic analyses further revealed a unique and heightened genomic footprint of selection when populations inhabited fragmented landscapes, most notably through an increased number of diversifying outliers. These unique genomic signatures were supported by a higher count of climate outliers in fragmented habitats compared to progeny originating from connected habitat. Interconnected habitats with a migration–selection–drift balance can suppress the divergence of weakly selected alleles due to continuous gene flow. Over time, this dynamic strengthens the persistence and linkage of alleles with larger selection coefficients, yielding fewer but more closely associated divergent alleles (Yeaman & Whitlock, 2011). In contrast, fragmented populations which a reduced gene flow can experience an enhanced fixation of beneficial alleles, even under weak selection, leading to more outliers (Blanquart et al., 2013; Legrand et al., 2017). Furthermore, the largely distinct adaptive signatures between fragmented and connected habitat, could point to altered climatic stressors in fragmented habitat. This phenomenon may be related to edge effects and diminished microclimatic buffering within fragmented habitats, potentially undermining established climate adaptations (Gehlhausen et al., 2000; Hofmeister et al., 2019). This aligns with findings that habitat fragmentation could disrupt trait-specific climate clines and reduce the efficiency of adaptive responses in *Primula elatior* to climatic stressors like drought (Van Daele et al. 2023). However, the capacity of species to adapt under rapidly changing conditions additionally depends on the diversity of adaptive alleles.

### 3.2. Habitat fragmentation and its impact on the climate-related adaptive potential

Habitat fragmentation did not significantly impact neutral genetic diversity in *P. elatior*, implying (i) a well-maintained drift–migration equilibrium (Lopez et al., 2009), or (ii) extinction debt. First, gene flow seems to be effective enough to counteract the potential loss of neutral genetic diversity due to drift. This observation aligns well with the one-migrant-per-generation rule, suggesting that even modest rates of migration—between one and 10 migrants per generation—can effectively mitigate the impacts of drift without impeding local adaptation (Mills & Allendorf, 1996). Second, an extinction debt in fragmented landscapes could forecast a future decline in populations, potentially causing an evolutionary delay wherein genetic diversity persists from historically connected habitats (Tilman et al., 1994; Vellend et al., 2006). If habitat fragmentation persists, the genetic diversity in *P. elatior* may further deteriorate, particularly if barriers to gene flow remain (Tigano & Friesen, 2016). As opposed to neutral genetic diversity, we observed a significant reduction of the adaptive genetic diversity as a consequence of habitat fragmentation. This could be attributed to the confounding effects of increased drift and selection, together accelerating fixation of loci under selection (Le Corre & Kremer, 2012). Although the reduction in adaptive potential may appear small, delayed evolutionary responses in addition to further climate change will likely result in significant losses in adaptive potential if conservation falls behind (Jump & Peñuelas, 2005). In addition, this finding suggests that outlier loci are better indicators of early genomic erosion in response to deteriorating habitat compared to neutral loci.

### 3.3. Floral traits in flux: herkogamy's alignment with climate and fragmentation

A significant proportion of outliers were significantly associated with herkogamy (N=242). These genetic variants were predominantly correlated with climate variables, specifically temperature and precipitation seasonality, in addition to moderate effects of habitat fragmentation and population size (Fig. 5). This finding aligns with the trait-based observation of Van Daele et al. (2023), who found that herkogamy and other floral traits of *P. elatior* showed considerable south-to-north variation with the anther–stigma distance decreasing significantly towards the north. The strong alignment of herkogamy to climate likely evolved during postglacial recolonisation of central and northern Europe. Many temperate forest species retracted to southern refugia during the last glacial period and postglacial recolonisation of dispersal limited species was generally dependent on rare medium- and long-distance dispersal events (Willner et al., 2023). These founder events often led to isolated populations at the northern range edge with increased levels of inbreeding and disrupted plant-pollinator effects (Jacquemyn et al., 2009; Willner et al., 2009). Such repeated founder effects can exert strong selective pressure on the S locus, culminating in the breakdown of self-incompatibility mechanisms (Yuan et al., 2023). This aligns with observations that reduced mate availability and inbreeding depression at range edges enhance the propagation and fixation of self-compatible variants (Encinas-Viso et al., 2020).

Similarly, the evolutionary pressures that lead to the erosion of herkogamy and self-incompatibility may also manifest in fragmented habitats, often as a consequence of diminished pollinator presence or a lower mate availability stemming from sparse plant density (Eckert et al., 2010). Even though the landscape history of habitat fragmentation displays large interregional variation across Europe, the increased land-use turnover is generally a more recent phenomenon (Kirby & Watkins, 1998). The reduced influence of habitat fragmentation on herkogamy may therefore mirror more recent evolutionary pressures, as indicated by the lower outlier count of SNPs related to both population size and habitat fragmentation. This aligns with the observation that a slight breakdown of herkogamy and self-incompatibility has been observed in fragmented populations, albeit smaller compared to the northwards reduction of herkogamy (Van Daele, Honnay, Janssens, & Kort, 2023). Consequently, habitat fragmentation and the resulting influence on the genetic variation underpinning herkogamy may have evolutionary implications for plant-pollinator interactions, potentially leading to further reduced outcrossing rates in *P. elatior*.

## 4. Conclusion

Where habitat fragmentation interferes with climate adaptation, it's vital that climate mitigation strategies are informed by the landscape context. While earlier studies predominantly concentrated on the general effects of fragmentation on genetic diversity, our research reveals that habitat fragmentation does not reduce the number of climate outliers, but triggers distinct evolutionary patterns in *Primula elatior*. We identified unique genomic markers of adaptation, an increase in adaptive outliers, and a diminished adjusted expected heterozygosity in diversifying outliers. To comprehensively understand the influence of habitat fragmentation on evolutionary processes and the adaptive capacity of plant species, it is thus imperative to consider both adaptive loci and neutral markers. Furthermore, we found that genetic variation associated with herkogamy reflected historical pollinator limitation driving mating system evolution. Our findings serve as an important catalyst for future research to explore the intricate interactions between habitat fragmentation, climate adaptation, and life-history traits in plant species. By uncovering historical and early signals of genomic erosion and mating system shifts, our study underscores the importance of considering the evolutionary impact of fragmentation across timescales. This insight is crucial for improving management strategies to maintain both genetic diversity and adaptive potential. Given that *Primula elatior* is a representative case for dispersal-limited forest herbs, it is reasonable to infer that our findings may be pertinent for other species with comparable ecological traits. Therefore, insights into the vulnerabilities and adaptive mechanisms of *Primula elatior* could provide a fundamental framework for shaping effective conservation approaches for other forest herbs with limited dispersal capabilities.